# GraphSense: Graph Embedding Based Code Suggestion Framework

H.R Navod Thisura Peiris
Department of Computer Engineering
*Faculty of Engineering*
University of Sri Jayewardenepura
Ratmalana, Sri Lanka
en97678@foe.sjp.ac.lk

*Abstract*— Code suggestions have become an integral part of IDEs and developers use code suggestions generated by IDEs all the time. These code suggestions are mostly for calling a method of an object or for using a function of a library and not for possible next line of the code. GPT based models are too slow or resource intensive for real-time code suggestions in local environments. As a solution to this GraphSense was introduced which provide code suggestions with minimum amount of resource usage in real-time.

## I. INTRODUCTION

Code suggestion systems like IntelliSense is built on to IDEs and only provides code suggestions based on method selection of a Class or Object and limited to line level suggestions. Recent additions like GitHub Copilot brings suggestions that even suggest next line on the fly but there is a limit to the usage in free-tier as actual model is cloud based. Local GPT based models give accurate suggestions but take considerable amount of resources and time. These limitations pose a research problem on how to develop and deploy an efficient code suggestion system locally. GraphSense is a framework introduced in this paper that solves these problems while reducing time and resource usage. GraphSense use a graph embedding model to suggest next line of code and achieved good improvement in inference time and memory usage while offering reasonable top-10 accuracy. Section II will describe related work. Section III will explain the underlying structure of framework and Section IV will show the results that was achieved.

## II. RELATED WORK

Embeddings are vector representations of knowledge such as images, text, audio. Embedding vector is a just a list of numerical values which represent some piece of information. Embeddings are widely used in AI realm to find similarity between text, images, audios etc. similarity is often calculated by utilizing Cosine distance or Euclidean distance between two embedding vectors.

### A. Code Representation and Embeddings

Effectively representing source code is crucial for tasks like code suggestion, completion, and search. Traditional methods rely on lexical and syntactic features, such as abstract syntax trees (ASTs) and control flow graphs (CFGs), to capture structural properties. More recent approaches use neural embeddings to convert code into meaningful vector representations. Techniques like Word2Vec, FastText, and Code2Vec learn code embeddings by analyzing syntax and semantics. Graph-based methods, such as Gated Graph Neural Networks (GGNNs), have shown better performance because they preserve relationships between program elements, making them ideal for code suggestion tasks.

### B. Graph-Based Representations in Code Analysis

Graphs naturally represent code structures by modeling relationships between variables, functions, and control dependencies. Graph-based models, such as AST-based Graph Neural Networks (GNNs), Code Property Graphs (CPGs), and Heterogeneous Graph Embeddings, capture contextual and semantic relationships more effectively than sequence-based models. Researchers have successfully used graph embeddings to improve tasks like bug detection, code summarization, and recommendation systems. By applying these techniques, code suggestion frameworks can understand long-range dependencies and contextual similarities between code snippets, leading to better recommendations.

### C. Code Completion and Suggestion Techniques

Researchers have explored several approaches to code completion and suggestion, including statistical models, rule-based techniques, and machine learning methods. Traditional models, such as n-grams and probabilistic context-free grammars (PCFGs), predict tokens based on previous code patterns but struggle to generalize. Deep learning methods, including LSTMs, Transformers, and BERT-based models like CodeBERT and GraphCodeBERT, have significantly improved code completion by learning rich contextual representations.

### D. Graph Embedding-Based Code Suggestion

Advances in graph embedding techniques have enabled more intelligent code suggestion frameworks. By modeling source code as graphs, these frameworks capture both local and global dependencies, improving the accuracy of next-token prediction, API recommendations, and function retrieval. Methods like GraphSAGE, Graph Attention Networks (GAT), and GGNNs generate high-quality code representations that generalize well across programming languages.

## III. METHODOLOGY

As introduced in Section I, GraphSense is focused on next code line suggestion with minimum amount of time and resource usage. GraphSense takes code files as a dataset and model the dataset as a graph where each node represents a code line and each edge represents a transition between them. Each edge has a weight which is the frequency of code line transition in the dataset. After modelling is done, framework creates an edge file with '.edg' extension which stores each unique edge in format: source node, destination node, weight. Now code is modelled as a graph and additionally edge file sharding is done to prevent Out-Of-Memory errors and training is done in batches. Graph is treated as undirected and reason is explained in sub-section B.

Model trained is a Word2Vec model which was trained on random walks generated over the graph. Random walks are just sequences of nodes in different traversals across the graph starting from each node. Walk length and number of walks can be set to any amount but for code completion task it can be as small as 10 walks of length 10 for each node. This is because code completion is more interested about closer code lines than further ones. Random walks are generated using pecanpy[7]. Pecanpy is a framework for fast and efficient random walk generation. Walks generated by pecanpy is then used to train the Word2Vec model. Gensim's[10] Word2Vec training API was used here.

### A. Hyperparameter tuning

Both simulating walks and model training require hyperparameter tuning. Pecanpy random walk simulation require setting hyperparameters *p* and *q*. *p* is the return parameter which controls the probability of revisiting the same node and *q* is the in-out parameter which controls the trade-off between BFS (breadth-first search) and DFS.

1) *p > 1: less likely to return to previous node*
2) *p < 1: more likely to return to the previous node*
3) *q > 1: BFS-like behavior*
4) *q < 1: DFS-like behavior*

Studies show that keeping *p* = 1 and *q* = 0.5 results in good clustering of nodes that interact with each other frequently.

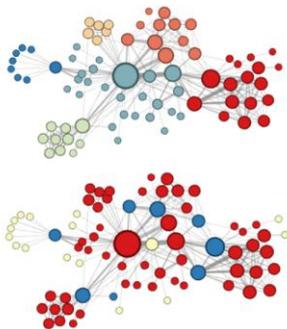

Figure 1: Clustering under different *p, q* values [3]

Figure 1(top) shows the clustering of nodes when *p* = 1 and *q* = 0.5. Figure 1(bottom) shows the clustering of nodes when p = 1 and *q* = 2. It is clear that using *p* = 1, *q* = 0.5 configuration identifies clusters of nodes that frequently interact with each other. [3]

Next hyperparameters are *num_walks* and *walk_length*. These are set as *num_walks* = 10 and *walk_length* = 10. This depends on the user's preference. For the experiments documented in this paper, the above values are used.

For Word2Vec model hyperparameters are *vector_size, window, min_count, workers, sg, hs, epochs*. *vector_size* is chosen as 128 to better represent distinct features. *window* is taken as 5 to consider next 5 code lines and previous 5 code lines in a random walk to generate embedding for current code line. This ensures that locality influences the embedding generation. *min_count* is taken as 1 to make sure all code lines which occurred at least one in the dataset were considered while training. *workers* are taken as {number of cpu cores} – 1 to make sure faster training. *sg* is set to 1 to make sure skip-gram architecture is used for better embedding rare code lines and *hs* is set to 1 to efficiently calculate approximate softmax value while using resources efficiently. *epochs* set as 100 to make sure embeddings are learnt well. Overfitting is not seen as a problem as this solution does not need generalization as embeddings are generated only using data seen in training corpus.

### B. Embedding Generation

Graph is actually treated as undirected to make sure that random walk does not abruptly stop when traversing code lines of a small code block. If there is only 3 code lines in a block. The random walk will stop from 3$^{rd}$ line if the graph is directed. If graph is undirected, random walk will continue to traverse backward and again forward until the *walk_length* is satisfied.

Consider a code block where only A and B are code lines random walk A, B, A, B, A, B, A, B, A, B will occur if it is an undirected graph on the other hand, if it is a directed graph random walk will be A, B. This allows algorithm to better generate embeddings as the idea is that if a code line appears in the same context frequently, its embedding will be closer to other code lines that share similar co-occurrences.

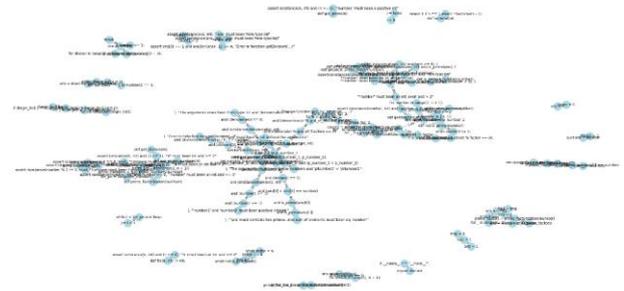

Figure 2: Code file modelled as a graph

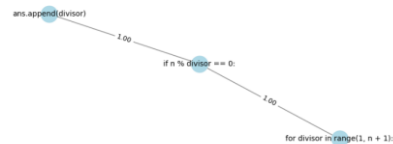

Figure 3: Code block modelled in a graph

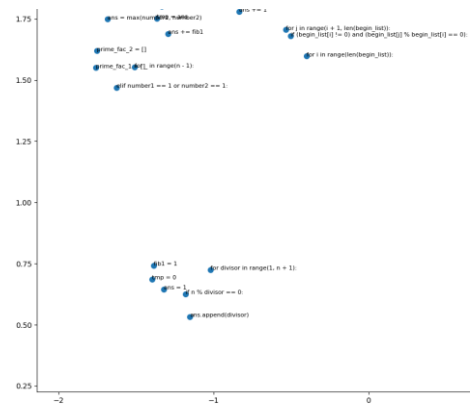

Figure 4: Vector space representation of code lines

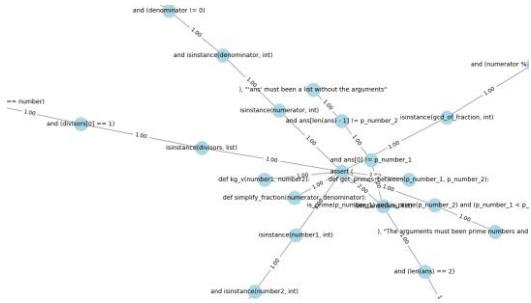
Figure 5: High degree node

Figure 2 shows a code file modelled as a graph there is a part which is highly connected. This is caused by lengthy code blocks or code lines that commonly appear in code blocks which binds subgraphs together. There are isolated subgraphs which are disconnected from large graph. They are the code blocks which has code lines that are only unique to those code blocks. Hence, they don't connect with any other subgraphs. Figure 3 shows a code block in graph and figure 4 clearly show that embedding vectors of those lines are placed close to each other. This makes similarity search accurate. However, high degree nodes are bit tricky as they are connecting to many subgraphs. Figure 5 shows a high degree node. If a code line that is related to a high degree node was used as current node to predict the next node, top 10 suggestions may not even contain correct result. This is a one drawback in this method. Comments were filtered out by the framework while preprocessing the dataset.

### C. Optimization

Optimization was needed as when the training dataset grows the model becomes larger in size due to growth of vocabulary. Solution was to save the embeddings of top frequent *N* code lines in a FAISS index and delete the model. For inference, the model is not needed as FAISS handle similarity search without the help of a model. *N* can range from thousands to millions. Casting embedding vectors from float32 to float16 reduced memory consumption from 50%. FAISS (Facebook AI Similarity Search [8]) is used for storage and efficient retrieval of vectors. FAISS was also the backbone behind the similarity search. FAISS index organizes the vectors in an index which reduce the search space of similarity search. FAISS provides $O(\log n)$ search time. There are multiple index types but in GraphSense, the IndexFlatL2 index type was used as it is the most accurate. IndexFlatL2 uses Euclidean distance between vectors to find closest vectors. The top 10 similar vectors are retrieved as top 10 code suggestions. As shown in figure 4, code lines which were seen closely in the dataset were placed closely in vector space making similarity search return the next possible line.

Deleting the model and storing vectors on FAISS has reduced space and memory usage significantly without harming accuracy of the solution but issue arise as code line to index mapping and index to code line mapping has to be stored separately. Simply storing these mappings as a serializable format such as joblib works but problem arise as the whole mapping of millions of items are loaded to memory. It consumes a large amount of memory. As a solution, code line to index and index to code line mappings were stored on RocksDB instances. RocksDB [9] is a high-performance, embeddable key-value store which can be embedded into file system without running a separate database server. RocksDB allow the retrieval of data only when needed without loading everything into memory. This further reduced the memory consumption. RocksDB takes less storage and memory as it is highly optimized.

### D. Handling OOV scenarios

Out-Of-Vocabulary (OOV) scenarios occur as real-life coding practices may differ from developer to developer and there can be many ways a code line can be written. As a solution for this, a separate FAISS index was maintained with text embeddings which were generated by a small embedding model called "all-MiniLM-L6-v2". Text embeddings were generated mainly by looking at text only. These embeddings were later used to handle OOV code lines.

First, the framework search existence of a vector for the code line provided. If not found, then the code line is an OOV code line. Then the "all-MiniLM-L6-v2" model was used to generate the text embedding and use FAISS index containing text embeddings to find most similar text embedding which correspond to the code line which is textually most similar to given code line. Model generates 384-dimensional embeddings and to convert them to 128-dimensional embeddings, a PCA model was trained and stored with other artifacts to be used in inference.

### E. Architecture and Implementation

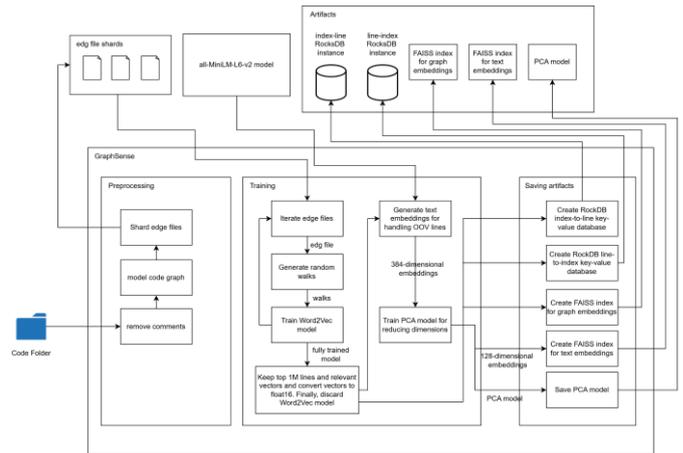
Figure 6: Training architecture

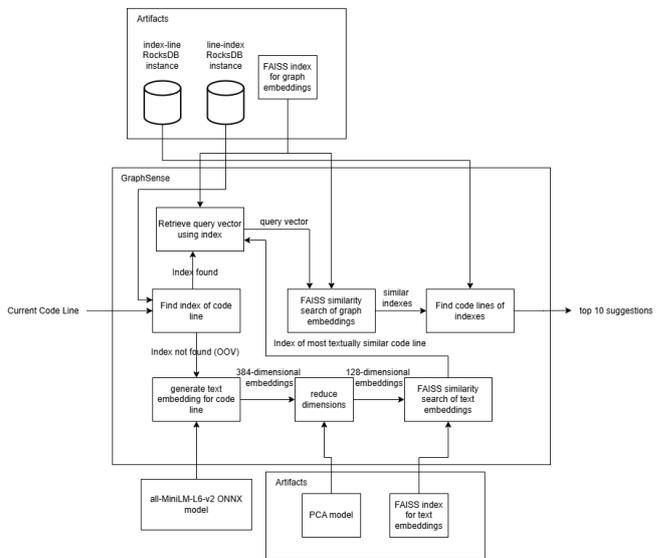
Figure 7: Inference architecture

Figure 6 shows the training architecture of the GraphSense framework. Code files were provided as input and graph modelling, training Word2Vec model, extracting embeddings to FAISS, storing code line to index and index to code line mappings on RocksDB instances are done inside the framework. 5 artifacts were generated and they are used in inference.

**Training Implementation**
1. Load code files
2. Preprocess code files
3. Shard edg files
4. for shard in shards
5.  generate random walks
6.  train Word2Vec model
7.  extract embeddings of top 1M code lines
8.  convert embeddings to float16
9.  extract code line to index and index to code line mappings
10. store vectors on main FAISS index and store mapping on RocksDB stores
11. generate text embeddings for vectors using all-MiniLM-L6-v2
12. convert text embeddings to 128-dimensional vectors using PCA
13. save PCA model
14. save text embeddings to text embeddings FAISS index

**Inference Implementation**
1. Get current code line
2. find index of code line using line to index mapping
3. if code line index is found
4.  retrieve related vector of code line for querying FAISS
5.  search main FAISS index for top 10 similar vector indexes
6.  use index to line mapping to find corresponding code lines
7. else
8.  generate text embedding using all-MiniLM-L6-v2
9.  convert to 128-dimensional vector using saved PCA model
10. use text embeddings FAISS index to find textually similar index
11. do steps 4 to 6
12. return top 10 suggestions (possible next lines)

## IV. EXPERIMENTS & RESULTS

Final Result of the framework is a ready to use solution for code suggestions. The framework was tested for its accuracy using Python version of TheAlgorithms dataset [11].

### A. Accuracy

TABLE 1: Top-k accuracies

| Dataset | Top-1 | Top-3 | Top-10 |
|---|---|---|---|
| TheAlgorithms | 0.4718 | 0.8012 | 0.8958 |

Top-1 accuracy is low as code lines which are previous to current line and code lines which are way ahead can even be situated closer than actual next line in the vector space. Top-3 and top-10 accuracies are high as next line can possibly fall within top 3 or top 10 suggestions.

### B. Performance and Scalability

Scalability of the framework shows linear results with no exponential growth in memory, space, time usage. As shown in figure 8, artifacts created with 0.1 million vocabulary took 61.3 MB and kept growing linearly and hit 561 MB for a 1 million vocabulary size. As shown in figure 9, memory occupied for inference for 0.1 million vocabulary took 273.77 MB and kept growing linearly and hit 734.58 MB for 1 million vocabulary. As shown in figure 10, Inference time taken by 0.1 million vocabulary took 0.0113 seconds and kept growing linearly and hit 0.0444 seconds for 1 million vocabulary.

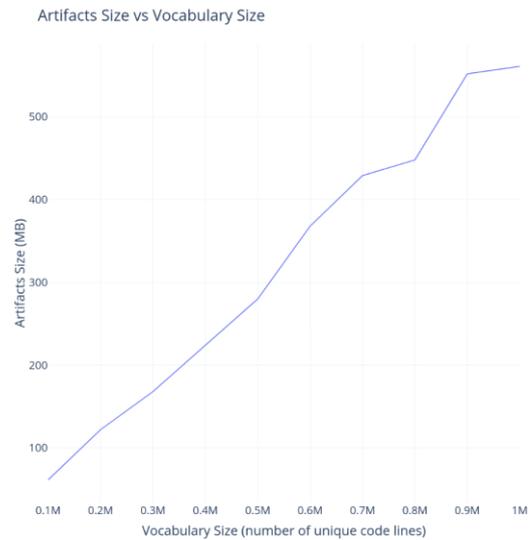
Figure 8: Artifacts size vs vocabulary size

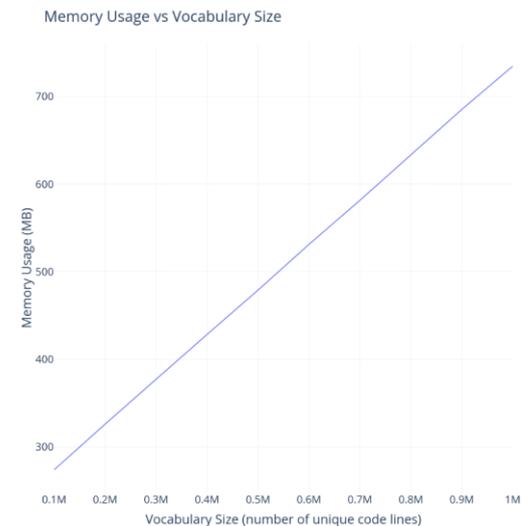
Figure 9: Memory usage vs vocabulary size

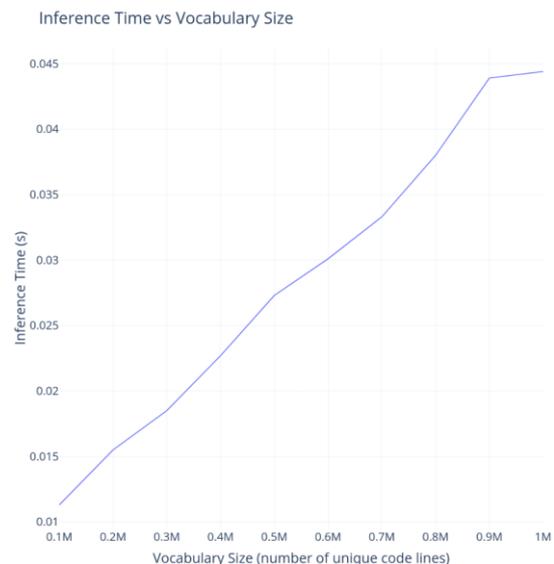
Figure 10: Inference time vs vocabulary size

These scalability values show that using GraphSense framework for up to 1 million vocabulary (unique code lines) is using less resources than most of the GPT-based models and is faster than GPT-based models in inference time. GPT2-Medium model uses 1.44GB storage and 1800MB memory during inference. It has inference time of 8 seconds on CPU and 2.26 seconds on RTX 4050 GPU.

Above tests for GraphSense were done while using CPU. Inference time shows significant improvement when compared to GPT2-Medium model with both CPU and GPU configurations.

## V. LIMITATIONS

Although performance and resource usage were good, there are some limitations to the framework.

*1) When vocabulary grows more than 2 million the memory usage becomes close to that of GPT2-Medium model.*

*2) Scaling to billion scale vocabulary is not possible on consumer grade computers.*

*3) Accuracy is less than that of GPT-based models.*

*4) Need to train for different programming languages seperately and artifacts has to be maintained seperately for each programming language.*

*5) High degree nodes may not get accurate suggestions as there are too many possible connections seen during training.*

## VI. DISCUSSION

The Problem this paper tried to solve was the resource intensive, expensive model usage in code suggestion tasks. As shown in section IV, GraphSense has achieved a reasonable accuracy given the nature of the method in use. It performed fast and efficiently considering a million-scale vocabulary of code lines. Inference is almost real-time. This provides a good ground for offline real-time code suggestions without relying on slower local models or expensive cloud-based models. Limitations are acknowledged in the section V. Furthermore, the community can build better frameworks on top of this with their own ideas for improvement. Improvements can be done such as using a better vector index type and using Approximate Nearest Neighbor (ANN) search but this can reduce accuracy in some cases. Memory usage grows with vocabulary size as FAISS loads vectors to memory on runtime. If a disk-based approach could be followed without compromising inference time, then this solution can scale better in terms of memory.

## VII. CONCLUSION

In conclusion, this paper shows that graph modelling based code suggestion systems can achieve reasonable accuracy and can be deployed in real-time environments to operate more efficiently than transformer-based models. Despite having drawbacks, the concept can further be utilized to create better versions as there is still room for improvements. Furthermore, GraphSense provides a framework for developing and deploying code suggestion solutions with minimal data preprocessing and resource usage. Users can leverage raw code repositories to train code suggestion models at ease. Lower training time and resource usage makes it more accessible and cheaper to train models and use the embeddings for code suggestions.